\documentclass[prb,aps,showpacs,twocolumn]{revtex4}
\usepackage{bm}
\usepackage{amsmath}
\usepackage{color}

\usepackage{graphicx}
\begin{document}

\title{Electrically controlled superconducting states at the heterointerface SrTiO$_3$/LaAlO$_3$}


\author{Keiji Yada}
\affiliation{Venture Business Laboratory, Nagoya University, Nagoya 464-8603, Japan}
\author{Seiichiro Onari}
\author{Yukio Tanaka}
\author{Jun-ichiro Inoue}
\affiliation{Department of Applied Physics, Nagoya University, Nagoya 464-8603, Japan}

\begin{abstract}
We study the symmetry of Cooper pair
in a two-dimensional Hubbard model with the Rashba-type spin-orbit interaction
as a minimal model of electron gas generated
at a heterointerface of SrTiO$_3$/LaAlO$_3$.
Solving the $\acute{\mbox E}$liashberg equation based on the third-order perturbation theory,
we find that the gap function consists of the mixing of the spin-singlet $d_{xy}$-wave component and the spin-triplet $(p_x\pm ip_y)$-wave one
due to the broken inversion symmetry originating from the Rashba-type spin-orbit interaction.
The ratio of the $d$-wave and the $p$-wave component continuously changes
with the carrier concentration.
We propose that the pairing symmetry is controlled by tuning the gate voltage.
\end{abstract}


\maketitle
Recent development of technology of epitaxial growth makes it possible
to fabricate heterointerface between two different transition-metal oxides.\cite{Ohtomo1,Okamoto}
The discovery of the generation of two-dimensional electron gas (2DEG) at
the $n$-type heterointerface SrTiO$_3$/LaAlO$_3$
attracts much interest\cite{Ohtomo} since both SrTiO$_3$ and LaAlO$_3$ are 
band insulators in the bulk material.
It is also noted that
this 2DEG shows superconductivity with transition temperature $T_{\rm c}=200$ mK.\cite{Reyren}
This superconductivity differs from the superconductivity of bulk SrTiO$_3$ with $T_{\rm c}=300$ mK in the carrier concentration and the dimensionality of mobile electrons.\cite{Suzuki}
The discovery of superconductivity is significant since it is promising to control the superconducting states electrically by the applied voltage.
Controlling the number of carrier by applied voltage in superconductors is one of the intriguing issues to understand the mechanism of the superconductivity.
In cuprates, a superconducting region is located next to an antiferromagnetic phase in an $n$-$T$ phase diagram.
However, in cuprate superconductors,
it is not easy to tune the number of carrier by external fields such as the magnetic field or the gate voltage
since the number of carrier depends on material compositions.
On the other hand, the carrier number of 2DEG at the heterointerface SrTiO$_3$/LaAlO$_3$
can be tuned by the gate voltage
since the carrier is generated by the polar and unpolar nature of LaAlO$_3$ and SrTiO$_3$, respectively.\cite{Nakagawa}
Actually, electrostatically tuned superconductor-metal-insulator transition was found to be accessible.\cite{Caviglia,Cen,Schneider}

When the electron correlations (Coulomb interaction) are responsible to the superconductivity,
the pairing symmetry depends on the number of carrier:
The spin-triplet $p$-wave pairing is favored at low density,\cite{Kohn,Chubukov,Nomura,Fukazawa,onari}
while on the other hand, the spin-singlet $d_{x^2-y^2}$-wave pairing is favored near half filling.\cite{Bickers,Moriya}
Thus, at the heterointerface of oxides with large Coulomb interaction,
the phase transition between the spin-singlet and spin-triplet states can be observed by the applied voltage.

Another characteristic feature at the heterointerface is lack of the inversion symmetry along the direction perpendicular to the interface ($z$-axis),
which induces the asymmetric spin-orbit interaction called the Rashba-type spin-orbit interaction (RSOI).\cite{Rashba}
This feature is common to non-centrosymmetric superconductors such as CePt$_3$Si, CeRhSi$_3$, and CeIrSi$_3$.\cite{Bauer,Kimura1,Sugitani}
In these superconductors,
the admixture of even- and odd-parity pairings has been suggested theoretically.\cite{fujimoto,yokoyama,yanase,tada}
Thus, we expect the mixing of parity also occurs in the present system.
In this case, the change in the pairing symmetry under applied voltage is not a phase transition
but a crossover between different parities.
Nevertheless, voltage control of the pairing symmetry in 2DEG generated at the oxides interface
could be a useful tool to study the carrier concentration dependence of the superconducting state.
Thus, it is very timely to reveal the pairing symmetry of Cooper pair
in 2DEG generated at the oxide interface.

In this Rapid Communication, we study the pairing state of superconductivity
at the heterointerface of SrTiO$_3$/LaAlO$_3$ by using the Hubbard model with the RSOI.
Since both SrTiO$_3$ and LaAlO$_3$ are band insulators in the bulk,
the number of carrier in 2DEG is far from the half filling.
In this case, the inherent strongly correlated nature of the titanium oxides may not emerge.
Therefore, we solve the $\acute{\mbox E}$liashberg equation based on perturbation theory up to the third-order expansion of the Coulomb potential.
We find that the pairing state in this system is the admixture of spin-singlet even-parity ($d_{xy}$-wave) pairing and
spin-triplet odd-parity ($p_{x}\pm ip_{y}$-wave) pairing,
which arises from broken inversion symmetry at the heterointerface.
The ratio of these two components continuously changes with the number of carrier.

The mobile electrons at the heterointerface are mainly introduced to $3d$ orbitals of Ti$^{3+}$ ions.
Considering the crystalline electric field by an ionic model,
the two-dimensional $d_{xy}$ orbital has the lowest orbital energy level in $3d$ orbitals.\cite{Maurice1}
In fact, it has been confirmed that the $d_{xy}$ electrons compose the 2DEG at the interface
by x-ray absorption spectroscopy with the linearly polarized light.\cite{Salluzzo}
Several results in density-functional calculations agree with this experimental one.\cite{Zoran,Pentcheva2}
Thus, we consider a two-dimensional Hubbard model with the RSOI as a minimal model of 2DEG generated at the heterointerface of SrTiO$_3$/LaAlO$_3$.
The Hamiltonian is given as
\begin{eqnarray}
H&=&-t\sum_{\langle i,j\rangle,\sigma}\left(c_{i\sigma}^\dag c_{j\sigma}+h.c.\right)
+\sum_{i}Un_{i\uparrow}n_{i\downarrow}
\nonumber\\&&{}-\frac{\lambda}{2}\sum_{k\sigma\sigma'}\left(\hat{\bm g}({\bm k})\cdot\hat{\bm \sigma}\right)_{\sigma\sigma'}c_{{\bm k}\sigma}^\dag c_{{\bm k}\sigma'},
\end{eqnarray}
where $c_{i\sigma}(c_{i\sigma}^\dag)$ is an annihilation (a creation) operator of an electron with spin $\sigma$ at site $i$, and $n_{i\sigma}=c_{i\sigma}^\dag c_{i\sigma}$.
$\langle i,j\rangle$ denotes the set of the nearest neighbor sites
and $t$ is the transfer integral.
The third term is the RSOI,
where $\lambda$ is the magnitude of Rashba field and $\hat{\bm \sigma}$ are the Pauli matrices.
The vector $\hat{\bm g}({\bm k})$, which satisfies the relation $\hat{\bm g}({\bm k})=-\hat{\bm g}(-{\bm k})$,
induces the breakdown of inversion symmetry.
We adopt
$\hat{\bm g}({\bm k})=(-v_y({\bm k}), v_x({\bm k}), 0)/t$,
with $v_{x,y}({\bm k})=\partial \varepsilon_{\bm k}/\partial k_{x,y}=2t\sin(k_{x,y})$.
The bare Green's function is given by the following $2\times2$ matrix in spin space,
\begin{eqnarray}
\hat G({\bm k},i\varepsilon_n)&\equiv&
\left(
\begin{array}{cc}
G_{\uparrow\uparrow}({\bm k},i\varepsilon_n)&G_{\uparrow\downarrow}({\bm k},i\varepsilon_n)\\
G_{\downarrow\uparrow}({\bm k},i\varepsilon_n)&G_{\downarrow\downarrow}({\bm k},i\varepsilon_n)
\end{array}
\right)\nonumber\\
&=&\left((i\varepsilon_n-\varepsilon_{\bm k})\hat {\bm I}+\frac{\lambda}{2}\hat{\bm g}({\bm k})\cdot\hat{\bm \sigma}\right)^{-1},
\end{eqnarray}
where $\hat {\bm I}$ is the unit matrix, and $\varepsilon_n=(2n-1)\pi T$ is the Matsubara frequency for fermions.

The effective pairing interaction $V_{\sigma_1\sigma_2}(k,k')$ in this study is given by the perturbation expansion up to the third order with respect to $U$,
\begin{align}
V_{\sigma_1\sigma_2}(k,k')=&V_{\sigma_1\sigma_2}^{\rm RPA}(k,k')+V_{\sigma_1\sigma_2}^{\rm Vertex}(k,k'),\\
V_{\sigma,-\sigma}^{\rm RPA}(k,k')=&U+U^2\chi(k+k')+U^3\chi^2(k+k')\nonumber\\
&{}+U^3\chi^2(k-k'),\label{vrpas}\\
V_{\sigma,-\sigma}^{\rm Vertex}(k,k')=&2U^3{\rm Re}\!\!\sum_q\! G(k-q)\chi(q)G(k'-q)\nonumber\\
&\!\!\!\!\!\!\!\!-2U^3{\rm Re}\!\!\sum_q\! G(k+q)\phi(q)G(-k'+q),\label{vvers}\\
V_{\sigma,\sigma}^{\rm RPA}(k,k')=&-U^2\chi(k-k'),\label{vrpat}\\
V_{\sigma,\sigma}^{\rm Vertex}(k,k')=&2U^3{\rm Re}\!\!\sum_q\! G(k+q)\chi(q)G(k'+q)\nonumber\\
&\!\!\!\!\!\!\!\!\!\!\!\!\!+2U^3{\rm Re}\!\!\sum_q\! G(-k+q)\phi(q)G(-k'+q),\label{vvert}
\end{align}
where $G(k)\equiv G_{\uparrow\uparrow}(k)=G_{\downarrow\downarrow}(k)$, $\chi(q)=-T/N\sum_k G(q+k)G(k)$, and $\phi(q)=-T/N\sum_k G(q-k)G(k)$.
$k\equiv({\bm k},(2n+1)\pi T)$ and $q\equiv({\bm q},2m\pi T)$ are the short notation of momentum and Matsubara frequency.
Here, we have dropped the terms including the off-diagonal part of $\hat G(k)$ for simplicity.
We divide $V_{\sigma_1\sigma_2}(k,k')$ into two parts $V_{\sigma_1\sigma_2}^{\rm RPA}(k,k')$ and $V_{\sigma_1\sigma_2}^{\rm Vertex}(k,k')$
for the later discussions.
$V_{\sigma_1\sigma_2}^{\rm RPA}(k,k')$ are the term included in the RPA, and $V_{\sigma_1\sigma_2}^{\rm Vertex}(k,k')$ are the contributions from the other terms.

In order to study the pairing instabilities,
we solve the linearized $\acute{\mbox E}$liashberg equation by a power method,
\begin{eqnarray}
\alpha\Delta_{\sigma_1\sigma_2}(k)&=&-\frac{T}{N}\sum_{k'}V_{\sigma_1\sigma_2}(k,k')F_{\sigma_1\sigma_2}(k'),\label{gap1}\\
F_{\sigma_1\sigma_2}(k)&=&\sum_{\sigma_3\sigma_4}G_{\sigma_1\sigma_3}(k)\Delta_{\sigma_3\sigma_4}(k)G_{\sigma_4\sigma_2}(-k).\label{gap2}
\end{eqnarray}
Solving Eqs. (\ref{gap1}) and (\ref{gap2}) self-consistently,
we obtain the eigenvalue $\alpha$ of the $\acute{\mbox E}$liashberg equation and pairing function $\Delta_{\sigma_1\sigma_2}(k)$.
The superconducting transition temperature 
$T_{\rm c}$ is identified as the temperature with $\alpha=1$.
Thus, the pairing symmetry with largest value of $\alpha$ is the most dominant.
Here, the indices of spin of the right-hand side and the left-hand side in Eq. (\ref{gap1}) are the same
since we have dropped the spin-flip scatterings in the effective interaction.
Therefore, the admixture of pairings is not induced by the effective interaction $V$.
However, the spin-flip processes in Eq. (\ref{gap2}) induce the admixture of pairings with different parities.

The spin-singlet and the spin-triplet components with $S_z=0$
are extracted by $\Delta_{\uparrow\downarrow}(k)\pm\Delta_{\downarrow\uparrow}(k)$
where $+1$ and $-1$ correspond to triplet and singlet ones, respectively.
The spin-triplet components with $S_z=\pm1$ are given by $\Delta_{\uparrow\uparrow}(k)$ and $\Delta_{\downarrow\downarrow}(k)$, respectively.
Due to the broken inversion symmetry, the components of odd-frequency gap function,
which are extracted by $\{\Delta_{\sigma_1\sigma_2}({\bm k},i\varepsilon_n)-\Delta_{\sigma_1\sigma_2}({\bm k},-i\varepsilon_n)\}/2$,
become finite in addition to even-frequency one.\cite{Yada}
However, we do not mention the odd-frequency component hereafter
since the magnitude of the odd-frequency component is much smaller than that of the even-frequency one.
We choose $t=1$ for the unit of energy.
We take $128\times128$ ${\bm k}$ meshes and 2048 Matsubara frequencies.

\begin{figure}[htbp]
\begin{center}
\includegraphics[width=7.5cm]{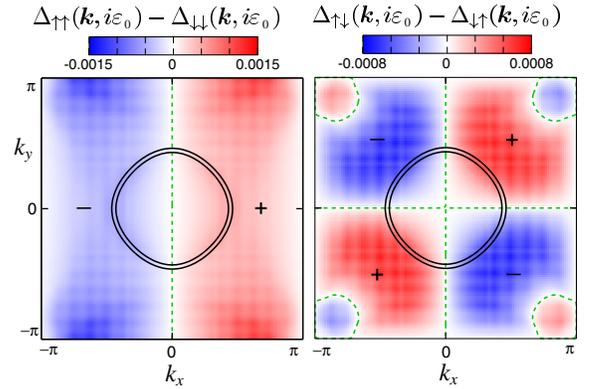}
\caption{(Color online) The ${\bm k}$-dependence of the spin-singlet gap function $\Delta_{\uparrow\downarrow}({\bm k}, i\varepsilon_0)-\Delta_{\downarrow\uparrow}({\bm k}, i\varepsilon_0)$ and the spin-triplet gap function $\Delta_{\uparrow\uparrow}({\bm k}, i\varepsilon_0)-\Delta_{\downarrow\downarrow}({\bm k}, i\varepsilon_0)$ for $T/t=0.008$, $\lambda/t=0.1$, $U/t=3$ and $n=0.3$,
with $\varepsilon_0=\pi T$.
The dotted (green) and solid (black) lines denote the nodal lines and Fermi surfaces, respectively.}\label{fig2}
\end{center}
\end{figure}

First, we look at the gap function.
In the absence of the RSOI, the spin state of Cooper pair can be classified into spin-singlet or spin-triplet states.
For $n\leq0.5$, the symmetry of the gap function with the largest values of $\alpha$ is $d_{xy}$-wave singlet or $p$-wave triplet pairing.
By introducing the RSOI, the pairing functions with different parity are mixed due to the breakdown of inversion symmetry,
since $G_{\sigma,-\sigma}(k)$ becomes finite in Eq. (\ref{gap2}).
For this reason, the $d_{xy}$-wave singlet pairing and the $p$-wave triplet pairing with $S_z=\pm1$ coexist.
Indeed, as shown in Fig. \ref{fig2}, we have confirmed that the pairing symmetry of the $\Delta_{\uparrow\downarrow}(k)-\Delta_{\downarrow\uparrow}(k)$ is $d_{xy}$ wave,
and that of $\Delta_{\uparrow\uparrow}({\bm k},i\varepsilon_n)$ and $\Delta_{\downarrow\downarrow}({\bm k},i\varepsilon_n)$ are $(p_x+ip_y)$- and $(-p_x+ip_y)$ waves, respectively.
Here, the component of $p$-wave triplet pairing with $S_z=0$ does not exist,
since the RSOI induces the mixing between the pairings with different parity of momentum and different values of $S_z$.\cite{Yada}

\begin{figure}[htbp]
\begin{center}
\includegraphics[width=6.0cm]{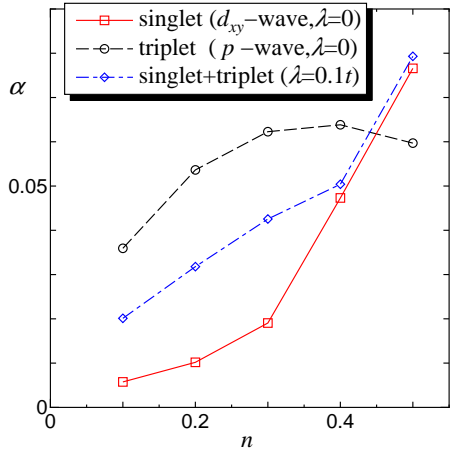}
\caption{(Color online) Eigenvalues of $\acute{\mbox E}$liashberg equation against $n$ for $T/t=0.008$ and $U/t=4$.}\label{fig1}
\end{center}
\end{figure}
Next, in Fig. \ref{fig1}, we show the $n$-dependence of $\alpha$ at $U/t=4$ and $T/t=0.008$.
At first, we concentrate on the case without the RSOI.
The value of $\alpha$ for the $d_{xy}$-wave pairing is the largest at $n\sim0.5$,
while on the other hand that for the $p$-wave pairing becomes the largest for $n\lesssim0.4$.
This result can be understood as follows.
Since $V_{\sigma,-\sigma}^{\rm RPA}(k,k')$ is the repulsive force ($V>0$),
$V_{\sigma,-\sigma}^{\rm RPA}(k,k')$ favors the singlet pairing
whose gap function changes the sign through ${\bm Q}$,
where ${\bm Q}$ is the wave vector at which the magnitude of $\chi({\bm Q}, i\omega_m)$ is large.
Since the magnitude of $V_{\sigma,-\sigma}^{\rm RPA}(k,k')$ increases with $n$,
the value of $\alpha$ for the $d_{xy}$-wave pairing also increases.
On the other hand,
the calculated $V_{\sigma_1\sigma_2}^{\rm Vertex}(k,k')$ is attractive for the $p$-wave pairing for $n\lesssim0.5$.
Since the relative contribution of $V_{\sigma_1\sigma_2}^{\rm Vertex}(k,k')$ in $V_{\sigma_1\sigma_2}(k,k')$ increases with decreasing $n$,
the $p$-wave pairing becomes dominant at low carrier concentration.
This is consistent with the previous studies using perturbative approaches including vertex terms.\cite{Kohn,Chubukov,Nomura,Fukazawa}
Notice that the values of $\alpha$ for $d_{x^2-y^2}$-wave pairing are smaller than those for $d_{xy}$-wave pairing for $n\lesssim0.5$
since $V_{\sigma,-\sigma}^{\rm Vertex}(k,k')$ suppresses it.
In the present calculation, the $d_{x^2-y^2}$-wave pairing becomes dominant for $n\gtrsim 0.6$ as noted in the previous studies in the context of high-$T_{\rm c}$ cuprates.\cite{Bickers,Moriya}

Next, we focus on the case of $\lambda/t=0.1$.
In the presence of the RSOI, breakdown of inversion symmetry induces the mixing of the even- ($d$-wave) and the odd-parity ($p$-wave) pairing.
For the small magnitude of $n$ ($n\lesssim0.4$), the main component of the gap function with largest $\alpha$ is the spin-triplet $p$-wave one.
This is because $V_{\sigma_1\sigma_2}(k,k')$ is attractive for $p$-wave pairing.
On the other hand, for $n\gtrsim0.4$, $V$ becomes attractive for $d_{xy}$-wave pairing and therefore,
the magnitude of the $d$-wave component becomes dominant.
Thus, the behavior of $n$-dependence of $\alpha$ is similar to that for the $d_{xy}$-wave and the $p$-wave pairings without the RSOI for $n\lesssim0.4$ and $n\gtrsim0.4$, respectively.

\begin{figure}[htbp]
\begin{center}
\includegraphics[width=6.0cm]{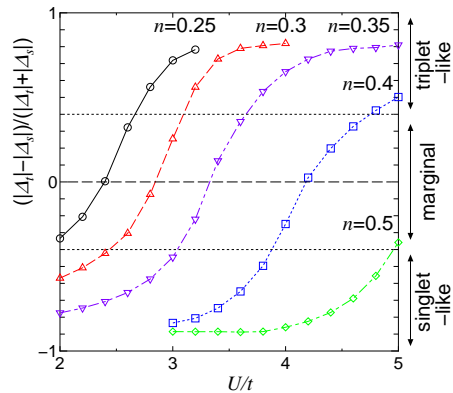}
\caption{(Color online) The ratio of the singlet and the triplet components for $T/t=0.008$ and $\lambda/t=0.1$ as a function of $U$.}\label{fig3}
\end{center}
\end{figure}

\begin{figure}[htbp]
\begin{center}
\includegraphics[width=6.0cm]{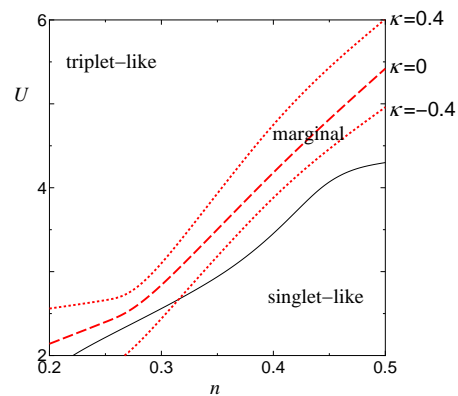}
\caption{(Color online) $n$-$U$ phase diagram of the symmetry of gap function at $T/t=0.008$.
The (red) dashed and (red) dotted lines denote the lines
where the corresponding values of $\kappa$ equal to $0$ and $\pm 0.4$, respectively.
The (black) solid line denotes the boundary between spin-singlet and spin-triplet phase in the absence of the RSOI.}\label{fig4}
\end{center}
\end{figure}

We have confirmed that the magnitude of the spin-singlet ($d_{xy}$-wave) and spin-triplet ($p$-wave) components changes with $n$.
Next, in Fig. \ref{fig3}, the ratio of singlet and triplet components
$\kappa=(|\Delta_t|-|\Delta_s|)/(|\Delta_t|+|\Delta_s|)$ is plotted as a function of $U$,
where $|\Delta_{s,t}|$ are defined as the absolute value of the maximum of the singlet and triplet components, respectively.
The values of $\kappa$ becomes unity for purely triplet case and minus unity for purely singlet case.
In the presence of the RSOI, $\kappa$ increases with $U$
since the magnitude of $V_{\sigma_1\sigma_2}^{\rm Vertex}(k,k')$ dominates over that of $V_{\sigma_1\sigma_2}^{\rm RPA}(k,k')$.
This is because the third-order term in $V_{\sigma_1\sigma_2}^{\rm Vertex}(k,k')$ is larger than that in $V_{\sigma_1\sigma_2}^{\rm RPA}(k,k')$ for $n\le 0.5$.
In the absence of the RSOI, the values of $\kappa$ change from $-1$ (pure singlet) to $1$ (pure triplet) abruptly.
On the other hand, the value of $\kappa$ varies continuously as a function of $n$ in the presence of the RSOI.

Here, we discuss the possibility of the crossover between the spin-singlet and the spin-triplet states
at the actual heterointerface SrTiO$_3$/LaAlO$_3$.
In Fig. \ref{fig4}, we show the $n$-$U$ phase diagram of the pairing states.
The dashed line denotes the line where the magnitude of the spin-singlet and the spin-triplet components are the same ($\kappa =0$).
This line corresponds to the dashed line in Fig. \ref{fig3}.
In both Figs. \ref{fig3} and \ref{fig4}, as a guide of eyes, we also draw the dotted lines with $\kappa=\pm0.4$,
where the corresponding values of the ratio $|\Delta_t|:|\Delta_s|$ are 7:3 and 3:7 for plus and minus sign, respectively.
We also depict the boundary between spin-singlet and spin-triplet phase in the absence of the RSOI
by the solid line.
In 3$d$ orbitals, the value of $U/t$ is popularly thought to be 3-4, which is approximately half of the band width.
In these values of $U/t$, the sign of $\kappa$ changes at around $n=0.3$-0.5.
If the number of carrier is controlled around these values by the applied gate voltage,
the crossover behavior between the spin-singlet and the spin-triplet states can be observed.
Note that $n=0.5$ is an ideal value for intrinsic doping, which originates from the polar and unpolar nature of LaAlO$_3$ and SrTiO$_3$, respectively.\cite{Nakagawa}
In order to observe this crossover, we need precise control of the carrier number and/or the large magnitude of the RSOI.
If the magnitude of the RSOI is small, the value of $\kappa$ varies abruptly with $n$ like the case without the RSOI.
Therefore, the crossover behavior in a small magnitude of the RSOI
is similar to a phase transition between the spin-singlet and spin-triplet states.
Since the qualitative results in the present calculation do not change with the values of $U$, $\lambda$ and $T$,
the crossover between the spin-singlet and spin-triplet states might be observed at the heterointerface SrTiO$_3$/LaAlO$_3$
or the related materials.
The direct determination of the superconducting energy gap from bulk property
is not easy since the transition temperature is very small in the present system.
One of the possible ways is to detect spin current via Andreev bound state.
We can expect the enhancement of the resulting spin current
when the magnitude of the triplet component of the pair potential
is larger than that of the singlet one.\cite{Vorontsov,Tanaka}

In this Rapid Communication, we have studied the pairing symmetry
in two-dimensional Hubbard model with the RSOI
considering the heterointerface of SrTiO$_3$/LaAlO$_3$.
Solving the $\acute{\mbox E}$liashberg equation based on the third-order perturbation theory,
we have found that the gap function consists of the mixing of the spin-singlet $d_{xy}$-wave component and the spin-triplet $(p_x\pm ip_y)$-wave one
because of the broken inversion symmetry originating from the RSOI.
The ratio of the $d$-wave and the $p$-wave components continuously changes
with the carrier concentration
through the change in the effective pairing interaction.


\begin{thebibliography}{99}
\bibitem{Ohtomo1}
A. Ohtomo, D. A. Muller, J. L. Grazul, and H. Y. Hwang,
Nature {\bf 419}, 378 (2002).

\bibitem{Okamoto}
S. Okamoto and A. J. Millis,
Nature {\bf 428}, 630 (2004).

%2DEG
\bibitem{Ohtomo}
A. Ohtomo and H. Y. Hwang,
Nature {\bf 427}, 423 (2004).

\bibitem{Reyren}
N. Reyren, S. Thiel, A. D. Caviglia, L. Fitting Kourkoutis, G. Hammerl, C. Richter,
C. W. Schneider, T. Kopp, A.-S. R$\ddot{\mbox u}$etschi, D. Jaccard, M. Gabay, D. A. Muller,
J.-M. Triscone, and J. Mannhart,
Science {\bf 317}, 1196 (2007).

\bibitem{Suzuki}
H. Suzuki, H. Bando, Y. Ootuka, I. H. Inoue, T. Yamamoto, K. Takahashi, and Y. Nishihara,
J. Phys. Soc. Jpn. {\bf 65}, 1529 (1996).

\bibitem{Nakagawa}
N. Nakagawa, H. Y. Hwang, and D. A. Muller,
Nature Mater. {\bf 5}, 204 (2006).

\bibitem{Caviglia}
A. D. Caviglia, S. Gariglio, N. Reyren, D. Jaccard, T. Schneider, M. Gabay, S. Thiel, G. Hammerl, J. Mannhart, and J.-M. Triscone,
Nature {\bf 456}, 624 (2008).


\bibitem{Cen}
C. Cen, S. Thiel, G. Hammerl, C. W. Schneider, K. E. Andersen, C. S. Hellberg,
J. Mannhart, and J. Levy,
Nature Mater. {\bf 7}, 298 (2008).

\bibitem{Schneider}
T. Schneider, A. D. Caviglia, S. Gariglio, N. Reyren, and J.-M. Triscone,
Phys. Rev. B {\bf 79}, 184502 (2009).

%3rd order
\bibitem{Kohn}
W. Kohn and J. H. Luttinger,
Phys. Rev. Lett. {\bf 15}, 524 (1965).
\bibitem{Chubukov}
A. V. Chubukov,
Phys. Rev. B {\bf 48}, 1097 (1993).
\bibitem{Nomura}
T. Nomura and K. Yamada,
J. Phys. Soc. Jpn. {\bf 71}, 1993 (2002).
\bibitem{Fukazawa}
H. Fukazawa and K. Yamada,
J. Phys. Soc. Jpn. {\bf 71}, 1541 (2002).


\bibitem{onari}
S. Onari, R. Arita, K. Kuroki, and H. Aoki,
Phys. Rev. B {\bf 73}, 014526 (2006).

\bibitem{Bickers}
N. E. Bickers, D. J. Scalapino, and S. R. White,
Phys. Rev. Lett. {\bf 62}, 961 (1989).

\bibitem{Moriya}
T. Moriya, Y. Takahashi, and K. Ueda,
J. Phys. Soc. Jpn. {\bf 59}, 2905 (1990).

\bibitem{Rashba}
E. I. Rashba,
Sov. Phys. Solid State {\bf 1}, 368 (1959).

%CePt_3Si
\bibitem{Bauer}
E. Bauer, G. Hilscher, H. Michor, Ch. Paul, E. W. Scheidt,
A. Gribanov, Yu. Seropegin, H. No$\ddot{\mbox{e}}$l, M. Sigrist, and P. Rogl,
Phys. Rev. Lett. {\bf 92}, 027003 (2004).

%CeRhSi_3
\bibitem{Kimura1}
N. Kimura, K. Ito, K. Saitoh, Y. Umeda, H. Aoki, and T. Terashima,
Phys. Rev. Lett. {\bf 95}, 247004 (2005).

%CeIrSi_3
\bibitem{Sugitani}
I. Sugitani, Y. Okuda, H. Shishido, T. Yamada, A. Thamizhavel, E. Yamamoto,
T. D. Matsuda, Y. Haga, T. Takeuchi, R. Settai, and Y. Onuki,
J. Phys. Soc. Jpn. {\bf 75}, 043703 (2006).

%rashba
\bibitem{fujimoto}
S. Fujimoto,
J. Phys. Soc. Jpn. {\bf 76}, 051008 (2007).
\bibitem{yokoyama}
T. Yokoyama, S. Onari, and Y. Tanaka,
Phys. Rev. B {\bf 75}, 172511 (2007).
\bibitem{yanase}
Y. Yanase and M. Sigrist,
J. Phys. Soc. Jpn. {\bf 77}, 124711 (2008).
\bibitem{tada}
Y. Tada, N. Kawakami, and S. Fujimoto,
New J. Phys. {\bf 11}, 055070 (2009).

\bibitem{Maurice1}
J.-L. Maurice, D. Imhoff, J.-P. Contour, and C. Colliex
Philos. Mag. {\bf 86}, 2127 (2006).

\bibitem{Salluzzo}
M. Salluzzo, J. C. Cezar, N. B. Brookes, V. Bisogni, G. M. De Luca,
C. Richter, S. Thiel, J. Mannhart, M. Huijben, A. Brinkman, G. Rijnders,
and G. Ghiringhelli,
Phys. Rev. Lett. {\bf 102}, 166804 (2009).

%band cal.
\bibitem{Zoran}
Z. S. Popovic, S. Satpathy, and R. M. Martin,
Phys. Rev. Lett. {\bf 101}, 256801 (2008).


\bibitem{Pentcheva2}
R. Pentcheva and W. E. Pickett,
Phys. Rev. B {\bf 78}, 205106 (2008).


\bibitem{Yada}
K. Yada, S. Onari, and Y. Tanaka,
Physica C {\bf 469}, 991 (2009).

\bibitem{Vorontsov}
A. B. Vorontsov, I. Vekhter, and M. Eschrig,
Phys. Rev. Lett. {\bf 101}, 127003 (2008).

\bibitem{Tanaka}
Y. Tanaka, T. Yokoyama, A. V. Balatsky, and N. Nagaosa,
Phys. Rev. B {\bf 79}, 060505(R) (2009).

\end{thebibliography}
\end{document}